# Usability and Security Effects of Code Examples on Crypto APIs

CryptoExamples: A platform for free, minimal, complete and secure crypto examples

Kai Mindermann
ISTE, University of Stuttgart
kai.mindermann@informatik.uni-stuttgart.de

Stefan Wagner
ISTE, University of Stuttgart
stefan.wagner@informatik.uni-stuttgart.de

*Abstract*—*Context*: Cryptographic APIs are said to be not usable and researchers suggest to add example code to the documentation. *Aim*: We wanted to create a free platform for cryptographic code examples that improves the usability and security of created applications by non security experts. *Method*: We created the open-source web platform *CryptoExamples* and conducted a controlled experiment where 58 students added symmetric encryption to a Java program. We then measured the usability and security. *Results*: The participants who used the platform were not only significantly more effective (+73 %) but also their code contained significantly less possible security vulnerabilities (-66 %). *Conclusions*: With *CryptoExamples* the gap between hard to change API documentation and the need for complete and secure code examples can be closed. Still, the platform needs more code examples.

## I. INTRODUCTION

Many cryptographic software libraries cannot be used easily and there are many approaches to improve (cryptographic) APIs [1]. A popular approach is not to improve the API itself but to develop tools that assist the developer using the APIs. The problem with tools is that, the programmer has to know about them, install them and use them. These 3 obstacles prevent many tools from being used. Another approach is to improve the APIs themselves. This is much harder, because often the API authors have to be involved. And even then it is difficult to change the API itself, because of backwards compatibility [2]. Improving the documentation is therefore a more feasible approach that also reaches the developers. One subset of the documentation are code examples. Many code examples can be found on the Q & A platform StackOverflow. Acar et al. [3] found out, that only 17% of the code examples are secure. Their conclusion is that "API documentation is secure but hard to use, while informal documentation such as Stack Overflow is more accessible but often leads to insecurity". As it can be difficult and a long process to get changes accepted for standard libraries [2], we describe our approach with *CryptoExamples* (www.cryptoexamples.com) and present empirical results.

## II. CRYPTOEXAMPLES

The goal of *CryptoExamples* is to offer *secure, complete, minimal, copyable, free and tested code examples* for common crypto scenarios in various programming languages. In the following we describe how these requirements and a few others are achieved:

1) The *common cryptographic scenarios* were derived from a study by Nadi et al. [4] on GitHub projects that apply cryptography. Most projects need or have problems with *symmetric/asymmetric encryption, signatures and verification*, and *generation and storage of secret keys*. 2) The platform is *open-source with an easy contribution and sanitation process*. It is easy to contribute new examples or update the existing ones, because examples are available in separate public repositories for each library. This means, that the crowd can not only report security problems, but also can provide solutions for them and they can easily be integrated. This reduces the time vulnerable code is available. 3) CryptoExamples tries to use *only standard library functionality to ensure copiability*. Examples should not introduce additional dependencies or require other configurations that could reduce their ease of integration. Therefore only APIs that are available in the programming language's standard library should be expected to be present for the example code. 4) We apply both static analysis tools and rely on expert code reviews to ensure *secure cryptographic functionality*. This is needed to keep up with new hardware generations that can apply brute force techniques faster and as new vulnerabilities are found, the parameters have to be kept up-to-date. 5) *Copyable code* is also made possible by using the free license "The Unlicense" that does not require developers to attribute or otherwise license the work. 6) The examples support the *latest stable releases of the programming language and related tools and libraries*. The examples should not cover all, especially not old, outdated and possibly insecure versions of programming languages. This backwards compatibility is currently not included in the platform but might be considered in the future. 7) Indicate *reviewed code*. The contributions to the example code repositories have to be reviewed by an expert for that domain and then proposed to the project maintainers for integration and wider publication via the website. Currently we follow a best effort principle. 8) We apply and require *automatic unit tests* for all code examples.

## III. EVALUATION

We performed a controlled experiment similar to the one by Mindermann, Keck, and Wagner [5] to find out how *CryptoExamples* changes the usability in practice. The task was to modify the application to not send plaintext but encrypt the text before with the Advanced Encryption Standard (AES). Half of the participants had to use *standard library of Java* with its Cryptography Extension (JCE) and the others a library which states that it focuses on usability, *Keyczar*. Then half of each group was given an opened web browser with the summary page for the corresponding library on *CryptoExamples*. This enabled them to look for a suitable example code that they could apply to the task in the experiment. Other participants could also use the web to search for example code, as all of them did. We made sure to check the screen recordings that none of these participants used *CryptoExamples*.

The 58 participants were all students with mostly software engineering as course of study (85%), with an average age of 21, and mostly male (2 unspecified, and 4 females). They rated their programming experience on average with $5.4 \in [0, 10]$ and their crypto and security knowledge on average with $3.9 \in [0, 10]$ (where 5 stood for average knowledge/experience).

We observed that the effectiveness (either the task was completed or not) for the task was increased on average by 73% with code example, also the Chi-squared test with no continuity correction suggests ($p = 0.035 < 0.05$) that we can reject the null hypothesis that there is an equal probability for success with or without example. If we look at both libraries individually we see that the biggest increase was for the Java Development Kit (JDK) (effectiveness increased by 184%).

Similar to the effectiveness, we observed an efficiency increase from 0.58 to 1.3 tasks per hour for participants that got an example. Wilcoxon's test suggests with $p = 0.008 < 0.05$ that we can reject the null hypothesis that the examples do not increase the efficiency. Interestingly, the efficiency also almost doubled for Keyczar. The examples make it significantly faster to implement crypto related tasks, at least for the scenario set in the experiment.

The examples also increased the satisfaction from 52.5 to 63.7 with example, measured with the System Usability Scale (SUS, Brooke [6]). This is also significant finding according to a t-test with $p = 0.01 < 0.05$ that suggests to reject the null hypothesis that the SUS is greater without example.

We analyzed the security of the produced code by counting warnings in the category *Security* (reported by *Find Security Bugs/SpotBugs*). We observed a drop by 66% for participants that used our examples. Here the Chi-squared test lets us reject the null hypothesis ($p = 0.037 < 0.05$) that there is no dependency between the number of security warnings and if the participants got an example.

## IV. Conclusion and future work

We wanted to close the research gap for cryptographic code examples [3] with the platform *CryptoExamples* that offers minimal, complete, tested and secure code examples. We think *CryptoExamples* is a big step in the right direction, because its goals and current state make it a good candidate to promote contributions to the documentation of APIs with appropriate and usable code examples. Also StackOverflow authors could link to these examples for reference, create new ones if the use case is common enough and even improve and sanitize existing examples with their knowledge on a central platform. The evaluation shows that the provided examples can improve several usability factors and increase the security significantly. Nevertheless, the examples still need to be fitting for a specific task, but we try to cover common scenarios [4]. Our project might become outdated too, but its simple contribution and sanitation concept based on git and GitHub make it a better alternative to the existing resources. An integration into the API documentation of the corresponding libraries would be ideal. For example Rust supports compiling code in the documentation, which makes it easier to maintain code examples together with the API. Now *CryptoExamples* has to be extended to provide examples for other programming languages, have better visualization of meta-data and experts have to be motivated to review the code examples. Additionally, we plan to extend the analysis of the evaluation results, e.g. analyze the answers to the Generic Cognitive Dimension Questionnaire for Evaluating Security APIs [7].

*Acknowledgments:* This work was partly funded by the Baden-Württemberg Stiftung.